\documentclass[aps,prb,twocolumn,superscriptaddress,showpacs]{revtex4}
\usepackage{epsfig,amssymb,amsfonts,amsmath}
\begin{document}


\title{Excess Modes and Enhanced Scattering in Rare-Earth Doped Amorphous Silicon Thin Films}


\author{B. L. Zink}
\email[]{bzink@boulder.nist.gov}
\affiliation{Department of Physics, University of California, San Diego\\ La Jolla, California 92093}
\affiliation{National Institute of Standards and Technology, 325 Broadway MC 817.03, Boulder, CO  80305}
\author{R. Islam}
\affiliation{Center for Solid State Science and Department of Physics and Astronomy, Arizona State University, Tempe, Arizona  85287}
\author{David J. Smith}
\affiliation{Center for Solid State Science and Department of Physics and Astronomy, Arizona State University, Tempe, Arizona  85287}
\author{F. Hellman}
\affiliation{Department of Physics, University of California, San Diego\\ La Jolla, California 92093}
\affiliation{Department of Physics, University of California\\ Berkeley, California  94720-7300}


\date{\today}

\begin{abstract}
We report specific heat and thermal conductivity of gadolinium- and yttrium-doped amorphous silicon thin films measured using silicon-nitride membrane-based microcalorimeters.  Addition of gadolinium or yttrium to the amorphous silicon network reduces the thermal conductivity over a wide temperature range while significantly increasing the specific heat.  This result indicates that a large number of non-propagating states are added to the vibrational spectrum that are most likely caused either by localized vibration of the dopant atom in a Si cage, or softening of the material forming the cage structures.  High-resolution cross-sectional electron micrographs reveal columnar features in Gd-doped material which do not appear in pure amorphous silicon.  Scattering from both the nanoscaled columns and the filled-cage structures play a role in the reduced thermal conductivity in the rare-earth doped amorphous semiconductor.  The overall result is an amorphous solid with a large bump in $C/T^{3}$ and no plateau in thermal conductivity. 
\end{abstract}


\maketitle

\section{Introduction}

A wide range of amorphous materials have similar features in specific heat, thermal conductivity and various spectroscopies that suggest a common physical origin.  The most notable of these features are a linear term in the specific heat, $C$, usually observed below $2$ K, thermal conductivity $k\propto T^{1.8}$ in the same temperature range,  a broad peak in $C/T^{3}$ between $10$ and $50$ K which is larger than in the corresponding crystalline phase, and a plateau in $k$ over the same temperature range.\cite{pohl}   The peak in $C/T^{3}$ is often correlated
with excess vibrational density of states seen in neutron or Raman scattering, referred to either as a ``boson peak"\cite{MalinovskySSC86} or ``excess modes."\cite{FabianPRL96,FeldmanPRB99} 
Despite several decades of study, no microscopic theory exists which offers a complete explanation of these phenomena. The standard tunneling model is a widely accepted description of the behavior
of $C$ and $k$ below $2$ K.  In this model the linear term in $C$ is attributed to a constant density of Schottky-like two-level systems (TLS), with the behavior of $k$ explained by scattering from these states.\cite{TLS,PhillipsJLTP1972}  This model offers little explanation of the physical nature or origin of the very closely spaced energy levels.   
There have been several efforts to develop physical models which also explain the $C/T^{3}$ peak and $k$ plateau.   Examples are the soft-potential model,\cite{BuchenauandRamos,BuchenauPRB1992} the fracton model,\cite{AlexanderPRB83,JagannathanPRB89} and a model relating dynamics at the glass transition to the low temperature phenomena.\cite{LubchenkoPNAS03,LubchenkoPRL01} 

Amorphous silicon ($a$-Si) 
forms a tetrahedrally bonded continuous random network and 
can be made only in thin-film form by evaporation, sputtering, and various chemical vapor deposition techniques.  Under-coordinated Si atoms form dangling bond defects, which can be reduced by introducing hydrogen either during deposition or after film growth.    The hydrogenated material, $a$-Si:H, shows improved electronic properties and has 
several important industrial applications for large area microelectronic devices. 
$a$-Si is often studied theoretically, in part 
because of its relatively simple structure, but also due to several predicted and observed deviations from typical amorphous solids.   
Both theory and experimental evidence suggest a strikingly low density of two-level systems, a relatively large change in the Debye temperature between crystalline and amorphous silicon, and a relatively small peak in $C/T^{3}$ in $a$-Si, deviating less from the Debye model than the crystalline phase.\cite{TLS,LiuPRL97,BergaSi,FeldmanPRB99,ZinkPRL06}   Properties ranging from electrical conductivity to density of TLS also show a strong dependence on the $a$-Si film growth method.\cite{PohlRMP02}  Another  example is the thermal conductivity plateau, though due to the difficulty of thermal measurements on thin-film samples, measurements of $k$ in the expected temperature range of the plateau exist for only three samples: an $\approx50$ 
$\mu$m thick sputtered sample reported by Pompe et al.\cite{Pompe} that shows a fairly well defined plateau, and much thinner $130$ and $277$ nm thick films recently reported by our group grown by e-beam evaporation that show no evidence of the plateau.\cite{ZinkPRL06}  
Both results can be reasonably explained by the theory presented by Feldman\cite{FeldmanPRB99} with different treatment of low-$Q$ scattering.       

Amorphous rare-earth-silicon alloys ($a$-RE$_{x}$Si$_{1-x}$, with RE=Gd, Y, Tb, etc.) prepared by e-beam co-deposition,
have shown many phenomena related to the interaction between 
local RE magnetic
moments and conduction electrons including a
very large negative magnetoresistance at low temperatures.\cite{prevpaper,pengpaper}
A negative Hall coefficient\cite{TeizerPRB2003} and thermopower\cite{ThermoPunpub} indicate that introducing the large, heavy RE adds carriers to the electron band, with the Gd ions contributing localized $S=7/2$ magnetic moments, while Y provides a non-magnetic counterpart with nearly identical ionic radius and valency. 
Recent computational and experimental 
results shed some light on the 
structure of $a$-Gd$_{x}$Si$_{1-x}$ and 
its non-magnetic analog, $a$-Y$_{x}$Si$_{1-x}$.  
Local density functional theory simulations of $a$-Y$_{x}$Si$_{1-x}$ suggest 
that Y$^{3+}$ ions are surrounded by low-coordinated Si, leaving them in a ``cage'' of dangling bonds.\cite{parrinello}
X-ray absorption fine structure (XAFS) experiments on $a$-Gd$_{x}$Si$_{1-x}$
using both the Si and Gd absorption edges
give a similar picture, with
Si atoms as nearest-neighbors to the Gd ions and no clustering.\cite{XAFS}
The XAFS studies also indicate that neither the Si coordination number
nor the Gd-Si distance changes with Gd composition, suggesting a Si cage but no
dangling bonds. Electron spin resonance (ESR) measurements indicate that the addition of even small amounts of Gd to the 
Si matrix eliminates the dangling bonds.\cite{OseroffESR}
Cages in $a$-Gd$_{x}$Si$_{1-x}$ are further supported by calculations using the 
full potential linearized augmented plane wave (FLAPW) method.\cite{WuFLAPW} 

This filled-cage structure is reminiscent of the filled skutterudite antimonides, such as CeFe$_{4}$Sb$_{12}$\cite{MorelliJAP95} or RE$_{1-y}$Fe$_{4-x}$Co$_{x}$Sb$_{12}$ (RE=La, Ce)\cite{SalesPRB97}, or Sr or Eu doped Ge clathrates,\cite{CohnPRL99} where the heavy La, Ce, Sr, or Eu dopant atoms 
fill cages in the host crystal structure.  Slack originally proposed that ``rattling" movement of the heavy dopant atom in the anharmonic potential of the cage would strongly scatter phonons but not charge carriers, resulting in a material with a high thermoelectric figure of merit.\cite{SlackCRCchap}  Although there is some debate about the detailed mechanism of the phonon scattering,\cite{FeldmanPRB2000,CaoPRB04} it is clear that the 
filled-cage structures in these materials 
dramatically alter the vibrational spectrum, reducing the thermal conductivity over a wide temperature range (at least $2-300$ K), and increasing the specific heat.\cite{KeppensNature98,FeldmanPRB2000,HermannPRL03}

In this paper we present specific heat and thermal conductivity measurements of gadolinium- and yttrium-doped amorphous silicon
thin films from $3-100$ K, as well as high-resolution cross-section transmission-electron microscope (XTEM) observations of the films.
We compare these measurements to data on $a$-Si films grown by the same technique and literature values for crystalline silicon to probe the nature of vibrational states and scattering in rare-earth doped amorphous silicon.    

\section{Experiment}

Thin film $a$-Y$_{x}$Si$_{1-x}$ and $a$-Gd$_{x}$Si$_{1-x}$ samples were e-beam co-evaporated from separate Y, Gd, and Si crucibles 
onto amorphous Si-N membrane-based microcalorimeters and amorphous Si-N coated Si substrates. 
The microcalorimeters and substrates were held near room temperature throughout the deposition, promoting the growth of amorphous films.   Typical deposition pressures were $\leq 2 \times 10^{-8}$ Torr.
The films on substrates were used to measure the film thickness via profilometry, the composition by Rutherford backscattering (RBS), the dangling bond density via ESR, the sound velocity,\cite{LeePRB05} and for XTEM imaging.  Both XAFS\cite{XAFS} and RBS measurements indicate atomic densities consistent with
pore-free films.  
A picosecond ultrasonic measurement\cite{LeePRB05} of  the longitudinal sound velocity in a $a$-Gd$_{x}$Si$_{1-x}$ film gave $v_{L}=(5.39 \pm 0.44) \times 10^{5}$ cm/s, lower than the value measured in our $a$-Si, 
$v_{L}=(7.51\pm 0.30) \times 10^{5}$ cm/s.   
Detailed description of the microcalorimetry techniques for measuring $C$ and $k$, including the determination
and subtraction of background contributions,
appear elsewhere.\cite{microcal,KappaPaper,RevazTCA05}

\section{Results and Discussion}

\begin{figure}
\epsfig{figure=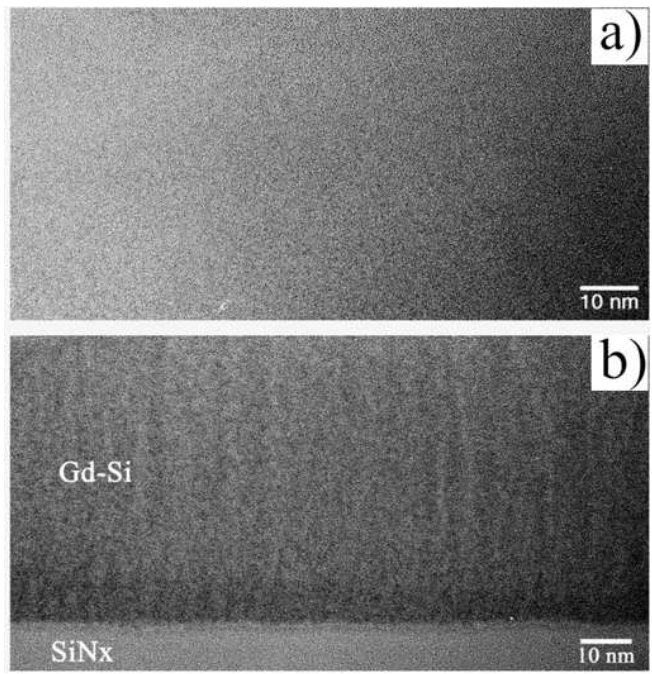,width=3.38in}
\caption{Low Magnification XTEM images comparing a) $a$-Si and b) $a$-Gd$_{18}$Si$_{82}$ thin films.  $a$-Si shows no evidence of crystallinity.  $a$-Gd$_{18}$Si$_{82}$ also shows no evidence of crystallinity, but does reveal columnar structures with several nanometer length scale.}
\label{TEMlow}
\end{figure}

\begin{figure}
\epsfig{figure=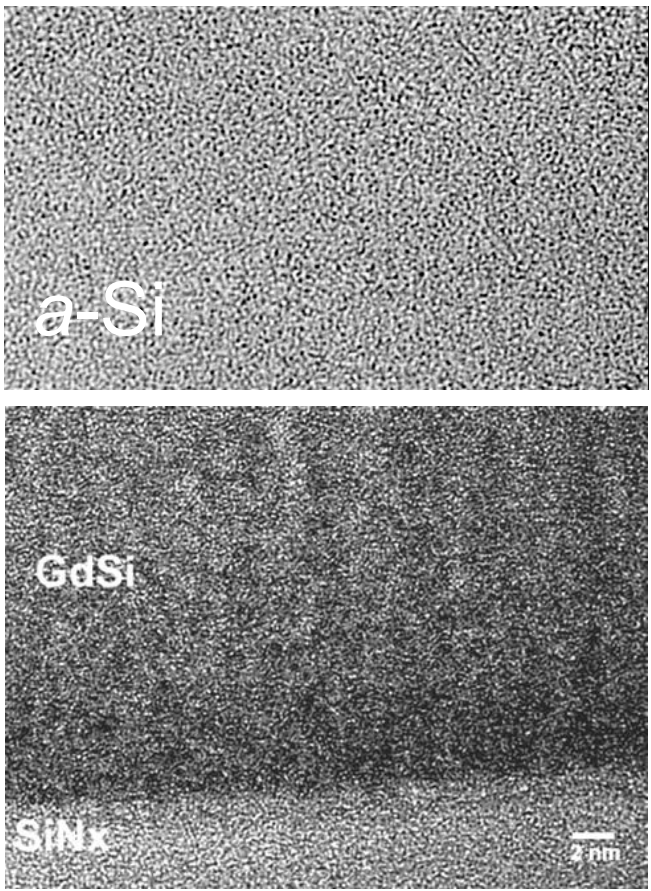,width=3.38in}
\caption{High resolution XTEM images of the same films clarify the results of Fig.\ \ref{TEMlow}.  
The approximate spacing of the columnar features in $a$-Gd$_{18}$Si$_{82}$ is $3-4$ nm.}
\label{TEMhigh}
\end{figure}

Figures \ref{TEMlow} and \ref{TEMhigh} compare XTEM images for $a$-Si and $a$-Gd$_{18}$Si$_{82}$ films.  The low magnification images (Fig.\ \ref{TEMlow}) show a featureless $a$-Si film, whereas the $a$-Gd$_{18}$Si$_{82}$ film grown under identical conditions shows a vertically streaked appearance indicative of a columnar structure.  
The higher magnification images in Fig.\ \ref{TEMhigh} also show featureless amorphous silicon, and  $3-4$ nm columnar features in amorphous gadolinium-silicon. 
Note that all images suggest dense, pore-free, clearly amorphous films at the atomic level.
Columnar microstructure, such as that seen in $a$-Gd$_{18}$Si$_{82}$, is a common outcome of the vapor deposition process for evaporated films when atomic mobility at the growth surface is low.

\begin{figure}
\epsfig{figure=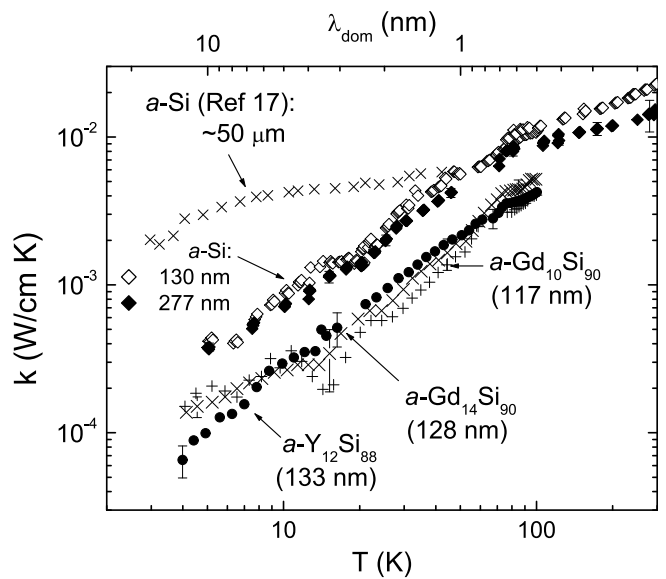,width=3.38in}
\caption{$k$ vs. $T$ plot comparing $a$-Y$_{12}$Si$_{88}$, $a$-Gd$_{10}$Si$_{90}$, and $a$-Gd$_{14}$Si$_{86}$ to thin e-beam evaporated $a$-Si films,\cite{ZinkPRL06} and previously published values for $a$-Si in the plateau regime.\cite{Pompe}}
\label{kcomp}
\end{figure}

Figure \ref{kcomp} compares $k$ of $a$-Y$_{x}$Si$_{1-x}$ and $a$-Gd$_{x}$Si$_{1-x}$ films to Pompe et al.'s sputtered $50$ $\mu$m thick $a$-Si film\cite{Pompe} and e-beam evaporated thin-film $a$-Si.  The top axis indicates the estimated wavelength of the vibrations which carry the most heat in the dominant phonon approximation,\cite{LambdaDomNoteReSi}  
calculated for $a$-Y$_{12}$Si$_{88}$. 
The e-beam $a$-Si shows a lower $k$ than the extremely thick sputtered film and no plateau.  Addition of Y or Gd to the material further reduces $k$ over the entire temperature range measured, and the three alloy films all have the same $k$ within error bars at all $T$. 
This reduction occurs despite the addition of electrons to the material, which increases the electrical conductivity dramatically.  However, this is not surprising, since a Wiedemann-Franz law estimation of the electronic contribution suggests that $k$ is totally dominated by vibrational excitations.\cite{BZthesis} 
In addition to the lack of composition dependence, $k$ of $a$-Gd$_{x}$Si$_{1-x}$ (which shows a huge negative magnetoresistance at low T) showed no measurable change in applied magnetic fields up to $8$ Tesla.

\begin{figure}
\epsfig{figure=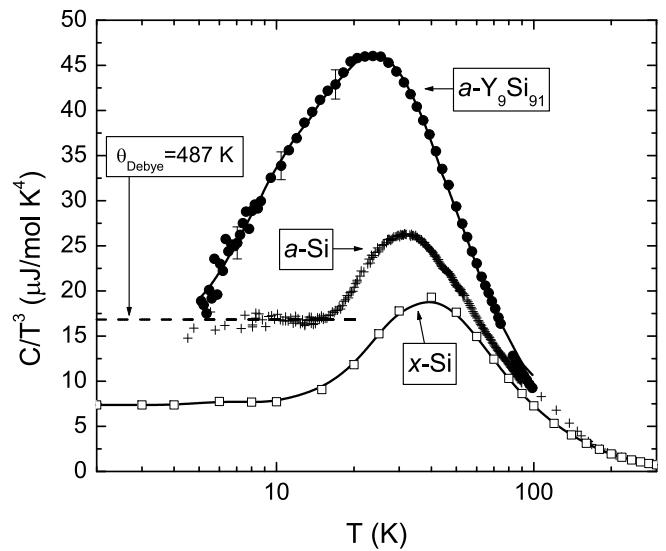, width=3.38in}
\caption{$C/T^{3}$ vs. $T$ plot showing $a$-Y$_{x}$Si$_{1-x}$, $a$-Si,\cite{ZinkPRL06} and $x$-Si.\cite{NBScp} The dashed line indicates the measured Debye contribution in $a$-Si, and the solid line through the $a$-Y$_{x}$Si$_{1-x}$ data is a fit described below.}
\label{CoverT3}
\end{figure}

Figure \ref{CoverT3} compares $C/T^{3}$ vs. $T$ (in J/g K$^{4}$) for $a$-Y$_{9}$Si$_{91}$  to thin-film $a$-Si\cite{ZinkPRL06} and bulk crystalline silicon.   The dashed line is the Debye specific heat function, $C_{D}$, for $\theta _{D}=487$ K.  
$C/T^{3}$ for $a$-Y$_{9}$Si$_{91}$ shows a much larger bump than that seen in either $a$-Si or crystalline Si. 
$a$-Y$_{x}$Si$_{1-x}$ for a broad range of $x$ have very similar specific heat
above $60$ $K$, and all show a large maxima in $C/T^{3}$ equal to or greater than that shown here. 
Samples with larger $x$ have electronic 
terms, $\gamma T$ (which appear as $\gamma/T^{2}$ on a $C/T^{3}$ plot), and 
an additional contribution to $C$ occurs in samples near the metal-insulator transition.\cite{BZthesis}
We have previously reported the specific heat of $a$-Gd$_{x}$Si$_{1-x}$, which is similar
to values for $a$-Y$_{x}$Si$_{1-x}$ above $60$ K, but is dominated at lower temperatures by 
large contributions from magnetic degrees of freedom.\cite{myPRL,Ternery,BZthesis}

\begin{figure}
\epsfig{figure=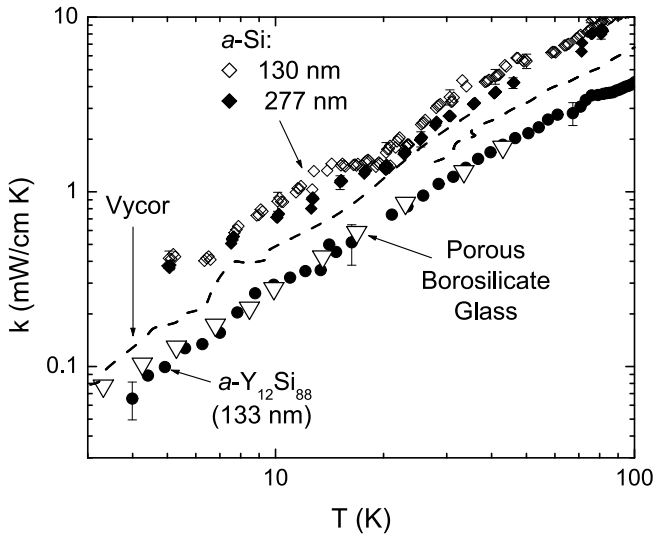,width=3.38in}
\caption{$k$ vs. $T$ for dense, pore-free $a$-Si and $a$-Y$_{x}$Si$_{1-x}$ thin films compared to data for Vycor, a silica glass with $29$\% porosity (dashed line),\cite{WatsonPRB03} and a borosilicate glass sample with artificially structured arrays of pores.\cite{ZaitlinPRB75} $k$ of $a$-SiO$_{2}$ is not shown but is only slightly higher than $a$-Si values at $10$ K.\cite{pohl}  At this temperature $k$ for these two porous materials is $2-4\times$ lower than for $a$-SiO${2}$, while adding Y to $a$-Si reduces $k$ by a factor of $\sim3$.}
\label{CompDep}
\end{figure}

Figure \ref{CompDep} compares $k$ vs $T$ below $100$ K for $a$-Si and $a$-Y$_{x}$Si$_{1-x}$ to two porous glasses, a porous silica (Vycor) sample with 29\% porosity\cite{WatsonPRB03} (dashed line), and a  glassy borosilicate material formed by fusing an array of capillaries ($\triangledown$).\cite{ZaitlinPRB75}  The data for both porous materials has been corrected for the missing volume of the pores.  Neither of these materials have a $k$ plateau in the typical temperature range, though Vycor shows an apparent plateau at much lower temperature.  
The reduction of $k$ in these porous materials is due to enhanced damping or scattering of long-wavelength modes caused by the pores.\cite{GracePRB86,ZaitlinPRB75}
The addition of Gd- and Y- to $a$-Si reduces $k$ by a quantitatively similar factor as does the addition of pores to $a$-SiO$_{2}$, but must rely on a different mechanism for scattering of heat carriers.     

\begin{figure}
\epsfig{figure=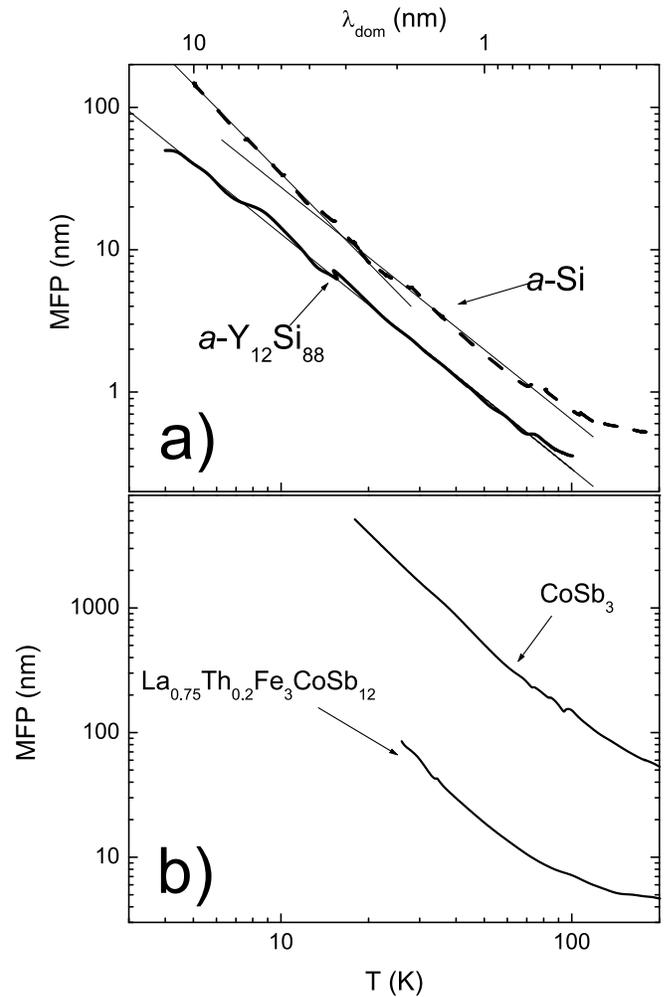,width=3.38in}
\caption{Estimated mean free path for vibrations vs. $T$.  a) $a$-Y$_{12}$Si$_{88}$ (heavy solid line) compared to a $a$-Si film (dashed line) grown by the same technique.  Solid lines indicate estimated $T$-dependence. Below $\sim 15$ K, where $\lambda_{dom}$ is $\approx$ to the spacing of the columnar features in the doped materials, the temperature dependence of the MFP in $a$-Si increases. b) Filled (La$_{0.75}$Th$_{0.2}$Fe$_{3}$CoSb$_{12}$) and unfilled (CoSb$_{3}$) skutterudite antimonides.\cite{SalesPRB97}  Filling the skutterudite cage structures with La reduces the estimated MFP over the whole measured temperature range.} 
\label{MFP}
\end{figure}

The structural evidence from XTEM (shown in Figs.\ \ref{TEMlow} and \ref{TEMhigh}), XAFS measurements, and simulations suggests two likely mechanisms for enhanced scattering of propagating vibrational excitations in $a$-RE$_{x}$Si$_{1-x}$ compared to $a$-Si: effects of the columnar structural features, and rattling of the caged Y or Gd dopants.   The measurements of $C$ and $k$ presented here indicate that both mechanisms play a role in the thermal properties of the rare-earth doped amorphous silicon.  
   
Figure \ref{MFP}a compares the estimated mean free path of vibrations for amorphous Si and $a$-Y$_{x}$Si$_{1-x}$ films.   This is given by $3k/C_{D}v_{D}$, where $v_{D}$ is the estimated Debye velocity and $C_{D}$ is the corresponding specific heat contribution from propagating modes, estimated here using $\theta_{D}=487$ K measured for $a$-Si.\cite{ZinkPRL06}     
In $a$-Y$_{12}$Si$_{88}$ MFP is roughly $\propto T^{-1.65}$ throughout the measured temperature range, while in $a$-Si the exponent increases below $\sim 15$ K  so that $MFP\propto T^{-2.08}$.  $15$ K is the temperature where $\lambda _{dom}$ is roughly equal to the spacing of the columnar features seen in  Figs.\ \ref{TEMlow}b and \ref{TEMhigh}b. Fig.\ \ref{MFP}b makes a similar comparison between filled and unfilled skutterudites.  The mean free path is again estimated using $3k/C_{D}v_{D}$, where $k$, $v_{D}$, and $C_{D}$ values reported
by Sales et al.\cite{SalesPRB97} Filling the cage in the skutterudite crystal with the heavy La dopant reduces the estimated mean free path over the whole measured temperature range.  
One interpretation of our data is that the overall suppression of $k$ in $a$-Y$_{x}$Si$_{1-x}$ is due to the effect of the rattling of caged rare-earth dopants, while the reduced temperature dependence of the MFP below $15$ K is the effect of the columnar structural features.  
This structural scattering need not be directly analogous to grain-boundary scattering in a crystal, but rather the effect of a softer region of the matrix occurring due to density fluctuations with the $3-4$ nm spacing seen in Figs.\ \ref{TEMlow}b and \ref{TEMhigh}b.

\begin{figure}
\epsfig{figure=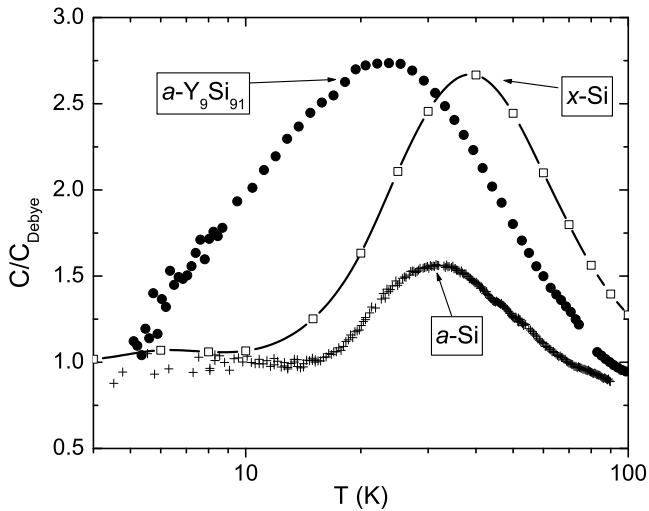,width=3.38in}
\caption{$C/C_{D}$ plot compares contributions to $C$ in excess of the Debye model 
for crystalline Si,\cite{NBScp} $a$-Si,\cite{ZinkPRL06} and $a$-Y$_{9}$Si$_{91}$. 
The low $C/C_{D}$ for $a$-Si compared to crystalline Si is atypical for amorphous systems, where the peak in $C/T^{3}$ is typically larger than in the corresponding crystal.  The larger $C/C_{D}$ in $a$-Y$_{9}$Si$_{91}$ is closer to expected behavior for amorphous systems.}
\label{Crat}
\end{figure}

Figure \ref{Crat} compares the ratio of the measured $C$ to the Debye contribution, $C_{D}$, for crystalline silicon, $a$-Si, and $a$-Y$_{9}$Si$_{91}$.  In typical amorphous solids such as $a$-SiO$_{2}$, $C$ is significantly larger than $C_{D}$, as a result of the excess modes.\cite{pohl}  We previously reported that the situation is very different in $a$-Si, where $C$ more closely matches the Debye model than does crystalline Si.\cite{ZinkPRL06} 
As shown in Fig.\ \ref{Crat}, 
the additional non-propagating vibrational states introduced by adding Y to the matrix 
cause $C/C_{D}$ for $a$-Y$_{9}$Si$_{91}$ to slightly exceed the value for crystalline Si, approaching the expectation for a typical amorphous material. 
It is also interesting to note that for $a$-Y$_{9}$Si$_{91}$, $a$-Y$_{13}$Si$_{87}$, and $a$-Y$_{21}$Si$_{79}$, the height of the peak in $C/T^{3}$, $P_{c}$, scales with its position in temperature, $T_{max}$, as $P_{c}\propto T_{\mathrm{max}}^{-1.6}$, which agrees with the scaling observed by Liu and Lohneysen for a wide range of amorphous solids.\cite{LiuEPL96}  Our recent $a$-Si measurement does not match this scaling behavior particularly well, indicating again that $a$-Si is a somewhat atypical amorphous material, while the additional non-propagating vibrational states cause $a$-Y$_{x}$Si$_{1-x}$ to show more typical behavior.  

The exact nature of these states is unknown, though it seems likely that excess modes are added due to 
locally softened Si cages around the heavy dopants or to anharmonic rattling of the dopant in the cage or both.
It is clear from the reduced $k$ that these excess modes do not carry heat.     
Several authors have reported a similar increase in the specific heat of filled skutterudites when compared to the ``empty" host crystal.  In the case of the skutterudites, the excess $C$ can be explained by the contribution of one or more Einstein modes, which each contribute a term $C_{\mathrm{E}}=(\theta_{\mathrm{E}}/T)^{2}e^{(\theta_{\mathrm{E}}/T)}/(e^{(\theta_{\mathrm{E}}/T)}-1)^{2}$, where $\theta_{\mathrm{E}}$ is the Einstein temperature.  The observation of these Einstein contributions provides direct evidence of the localized ``rattling" of the filler atom, and is corroborated by inelastic neutron scattering data, resonant ultrasound spectroscopy, and simulations.\cite{KeppensNature98,FeldmanPRB2000,HermannPRL03}
Following the ``rattling" atom analogy, we fit the measured specific heat of $a$-Y$_{9}$Si$_{91}$ to the equation $C=C($$a$-Si$)+A_{1} C_{\mathrm{E1}}+A_{2} C_{\mathrm{E2}} + A_{3} C_{\mathrm{E3}}$.  A fit of this type assumes that the filling of the cages has little effect on the elastic properties of the host material.  Though we believe this is true to good approximation for the $a$-Y$_{9}$Si$_{91}$ sample,  available evidence suggests a possible reduction in $\theta_{\mathrm{D}}$ for more heavily Y-doped $a$-Si.\cite{BZthesis}  This type of effect can have important implications for the physical interpretation of the resulting fit parameters.\cite{FeldmanPRB2000}   The solid line in Fig.\ \ref{CoverT3} represents the fit with $A_{1}=1.0$ J/mol K, $\theta_{\mathrm{E1}}=194$ K, $A_{2}=0.83$ J/mol K, $\theta_{\mathrm{E2}}=100$ K, and $A_{3}=0.086$ J/mol K, $\theta_{\mathrm{E3}}=52$ K.  The data can be fit somewhat less well by two Einstein modes ($\sim108$ and $51$ K) , but is very poorly modeled with a single Einstein contribution.\cite{aSi02fitnote}  As a comparison, Keppens, et al. used Einstein modes at $70$ and $200$ K to explain the contribution of the rattling La atom in La$_{0.9}$Fe$_{3}$Co$_{}$Sb$_{12}$,\cite{KeppensNature98} while Hermann, et al. needed only a single Einstein mode at $53$ K to describe thallium rattling in similar antimony skutterudites.\cite{HermannPRL03} 
The similarity of these results to the Y-doped amorphous Si data presented here suggests that similar physics drives the excess modes and reduced $k$ in these two classes of materials.   

Regardless of the exact nature of the contributions to $C$ in $a$-Y$_{x}$Si$_{1-x}$, the combined effect of the addition of excess modes and the enhanced scattering is an amorphous material with a large bump in $C/T^{3}$ but no corresponding plateau in $k$.  We conclude that these two phenomena, which are often both observed in a given amorphous material, are not necessarily the work of the same physical mechanism.  
This suggests an interesting future experiment.  It is currently a matter of debate whether the $k$ plateau and $C/T^{3}$ bump and the related excess modes can be explained within a single theoretical framework with the TLS that dominate the low $T$ properties of most amorphous insulators.  Our work shows that addition of heavy Y dopants to $a$-Si, a material which has few excess modes and very low contributions from TLS, adds excess modes to the material.  Study of $a$-Y$_{x}$Si$_{1-x}$ below $2$ K could indicate whether TLS have returned with the addition of Y, which would provide evidence of correlation between TLS and excess modes.   
This also suggests that addition of heavy dopants could provide a potentially tunable method for adding excess modes to $a$-Si, allowing systematic study of the vibrational excitations in amorphous insulators.   

\section{Conclusions}

In summary, we measured specific heat and thermal conductivity of Gd- and Y-doped amorphous Si thin films.  Addition of the heavy dopant atoms introduces scattering which reduces the thermal conductivity below values measured for pure $a$-Si.  At the same time, a large number of excess vibrational modes are added, resulting in a bump in $C/T^{3}$ which is much larger than that in the pure amorphous material and also larger than in crystalline Si, in better agreement with expected behavior of $C$ in amorphous insulators.  This bump can  similar manner as the contribution of rattling modes to filled antimony skutterudites, suggesting that similar physics drives the thermal behavior of these rather different systems.   Furthermore, the ability to add excess modes to $a$-Si suggests that continuing study of heavy-atom doped amorphous silicon could enable systematic probe of the correlation between tunneling systems and excess modes in amorphous materials. 

\section{Acknowledgments}
We thank B. Maranville, M. Liu and M. Wong for the RBS measurements,  S. Oseroff and C. Rettori for the ESR measurement,
D. Cahill for the sound velocity measurement and many helpful comments, B. Pohl, A. Migliori and R. Dynes for fruitful
discussions, and the NSF and LANL for support.  We also acknowledge use of facilities in the John M. Cowley Center for High Resolution Electron Microscopy at Arizona State University.



\begin{thebibliography}{47}
\bibitem[{aSi()}]{aSi02fitnote}
\bibinfo{note}{Even after allowing the linear contribution from the TLS, a fit
  of this type does not explain Zeller and Pohl's measured specific heat of
  $a$-SiO$_{2}$ well. This is in line with the view that at least some of the
  excess modes in typical amorphous materials carry heat, and that these added
  heat carriers contribute to the $k$ plateau.}
\end{thebibliography}
\end{document}